# Bridging the Skills Gap: Evaluating an AI-Assisted Provider Platform to Support Care Providers with Empathetic Delivery of Protocolized Therapy

**William R Kearns, Ph.D.[1], Jessica Bertram, MS[1], Myra Divina, MS, BS[2], Lauren Kemp MN, RN[2], Yinzhou Wang, BA[1], Alex Marin, Ph.D.[1,3], Trevor Cohen, MBChB, Ph.D.[1], Weichao Yuwen, Ph.D., RN[2]**

[1]University of Washington, Seattle, WA; [2]University of Washington, Tacoma, WA; [3]Microsoft, Redmond, WA

**Abstract**

*Despite the high prevalence and burden of mental health conditions, there is a global shortage of mental health providers. Artificial Intelligence (AI) methods have been proposed as a way to address this shortage, by supporting providers with less extensive training as they deliver care. To this end, we developed the AI-Assisted Provider Platform (A2P2), a text-based virtual therapy interface that includes a response suggestion feature, which supports providers in delivering protocolized therapies empathetically. We studied providers with and without expertise in mental health treatment delivering a therapy session using the platform with (intervention) and without (control) AI-assistance features. Upon evaluation, the AI-assisted system significantly decreased response times by 29.34% (p=0.002), tripled empathic response accuracy (p=0.0001), and increased goal recommendation accuracy by 66.67% (p=0.001) across both user groups compared to the control. Both groups rated the system as having excellent usability.*

**Introduction**

Mental health conditions are highly prevalent and exert a considerable burden on society, with a global estimated cost of 125.3 million disability-adjusted life years in 2019[1]. Unfortunately, despite this high prevalence and societal cost, services for these conditions remain in short supply[2,3]. This unmet need for mental health services has shifted to care providers in primary care, nursing, coaching, or peers who have not been extensively trained to provide mental health support[4]. Consequently, recent work has explored the use of artificial intelligence (AI) methods to support peer-to-peer supporters without mental health training[5,6]. On account of its role as a predictor of therapy outcomes[7], a particular focus of this work has been on *empathy*, which from a psychotherapeutic perspective concerns the ability to establish rapport, and attend to and understand the communication of the emotions and experiences of others[7,8]. Sharma et al. describe a system (HAILEY) that suggests edits to peer-to-peer support messages with positive results of increased empathy expressed by the peer supporters[6]. However, it is not apparent whether such a system would increase either the empathy expressed by care providers, or their efficiency, given they would first have to craft the message before being provided feedback which they then need to read and decide whether to incorporate. The utility of empathy-related AI support for provider *selection* of professionally crafted text messages remains unevaluated and is the focus of the current work. Response retrieval has several benefits over generation including the safety and controllability of the responses.

This work presents a step toward addressing the supply-demand imbalance between those who need mental health services and those who can provide treatment, by both applying conversational AI to reduce response times during asynchronous communication for more immediate engagement and also lowering the educational barrier to entry, thereby increasing the number of individuals who can provide care. A particular focus of A2P2 was support for the selection of *empathic responses* - dialog responses that demonstrate an understanding of the client's situation, perspective, and feelings. We provide a preliminary evaluation of the potential for this technology to achieve these goals by measuring the speed and accuracy of care providers with and without expertise in mental healthcare in responding to messages in mock therapy sessions. The aims are: 1) to evaluate the preliminary efficacy of the provider platform in improving the empathy, efficiency, and treatment accuracy of provider messages, and 2) to describe the usability and feasibility of the platform for use in the delivery of protocolized therapy.

## Methods

The study was approved by the University's Institutional Review Board and usability testing was conducted at the start of 2022. Building upon learnings from usability testing of a Wizard of Oz (WOZ) system[9] used for human-in-the-loop conversational AI development with standardized patients to deliver a protocolized therapy called Problem-Solving Therapy (PST) for family caregivers to manage their caregiving symptoms such as stress and anxiety, A2P2 was redesigned into a provider-centric interface to improve the efficiency of communication with clients to collect data more efficiently while simultaneously providing care.

### System Description

A constrained version of the provider platform was used in this study to focus the evaluation on the effect of a response suggestion system (Figure 1). The study team designed a PST protocol to address common caregiving *symptoms* (e.g., *fatigue*, *disturbed sleep*, *depressive symptoms*, and *anxiety*) through a conversational AI system that supports caregivers in monitoring their symptoms, developing problem-solving skills, and taking action toward their goals over 4-5 sessions[9]. The first session focuses on identifying a *symptom* and its associated problems, and developing a potential *solution* to be tried out for the following week. The remaining sessions follow a similar pattern, evaluating the *solution* and co-developing new *solutions* if necessary. At each turn of the session, the provider is prompted to provide both an empathetic response that acknowledges the client's message and therapeutic response that advances the dialog through the steps of PST.

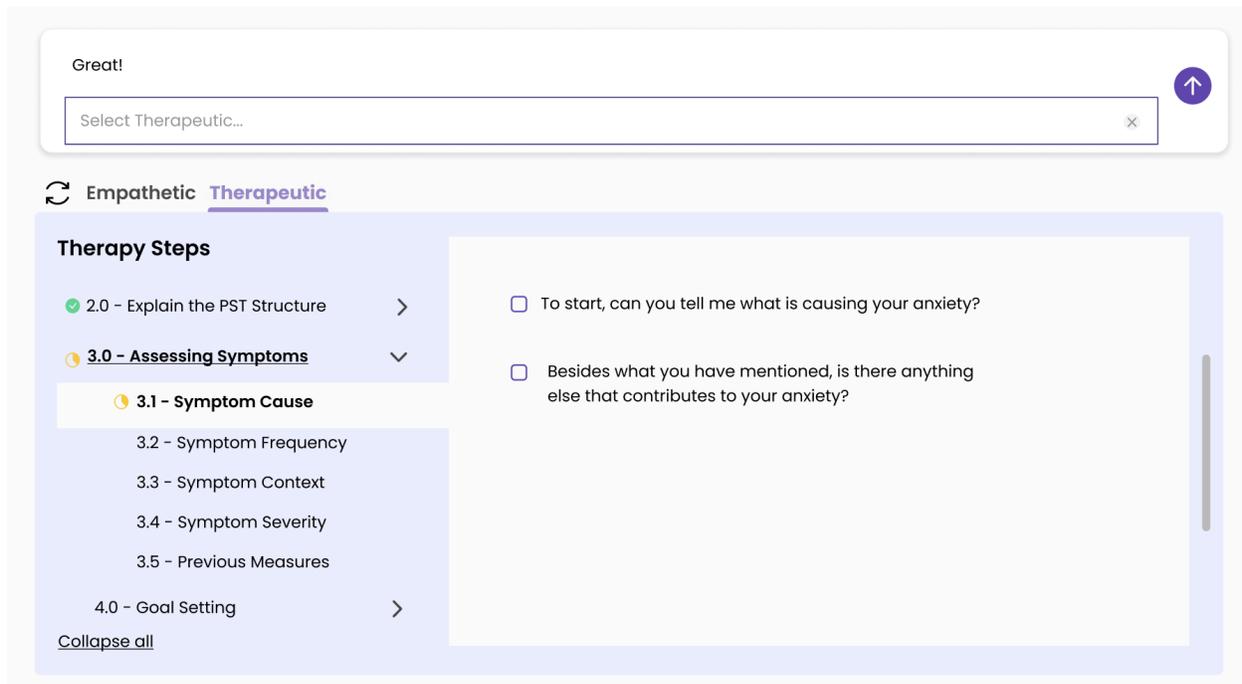

**Figure 1:** Conversational assistance frame within the provider platform showing the therapeutic response suggestions (right) and PST step tracker (left).

The core of the provider platform is a knowledge-based conversational AI system designed to guide a user in providing mental health support using natural language understanding (NLU), dialog management (DM), clinical knowledge graph (CKG), and natural language generation (NLG) components. The system architecture is as follows:

1) The NLU component processes client messages received by A2P2 using a Rasa[10] NLU pipeline trained to recognize mental health *symptoms*. If a *symptom* is detected, it is stored as a *slot* in a conversational state tracker. Additionally, the NLU component infers the *emotional state* of the client.

2) The DM consists of two parts, a conversational state tracker and a finite-state policy based on the PST steps. At the start of a session, the conversational state tracker is initiated with information from the client profile including their *name* and any existing *symptoms*, *goals*, or *solutions*. As the session progresses, the state tracker stores new information detected by the NLU component and persists this information back to the knowledge graph to enable the recommendation of *goals* and *solutions* that help with and address the client's latest *symptom*.

   A finite-state policy is used to suggest therapeutic responses based on the session number and this component is visually surfaced to care providers in the form of a checklist referred to as the PST step tracker (left side of Figure 1) to support them in orienting themselves to where they were in a session and facilitate them completing all the steps of PST. A check mark serves as a visual indicator of which steps have been completed and which are remaining. At any time, the care provider can select a therapy step from the PST step tracker to bring up all responses for that step. The interface sends requests for the currently selected step within the checklist.

3) The CKG was constructed by subject matter experts using a schema based on the PST process. This is used to suggest contextually-aware combinations of *goals*, *solutions*, and *resources* to a client based on their *symptoms* (Figure 2). In total, the knowledge graph contains 23 *goals* with 12 caregiving *symptoms* and 21 *solutions* which each have one associated *resource*. A *solution* can help with more than one *goal,* and there are 56 connections between these two node types. Similarly, a *goal* can address more than one *symptom,* so there are 119 connections between these two node types. The platform automatically links clients to their symptoms, goals, and solutions as the care provider completes PST steps. The CKG provides persistent storage for the client's session history to provide up-to-date recommendations based on the client's needs and actions.

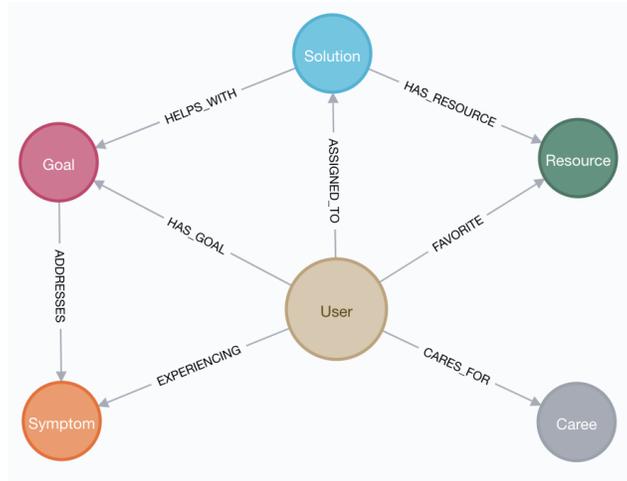

**Figure 2:** Schema of the Clinical Knowledge Graph used to store user data for predictions.

4) The NLG system has a one to many mapping between DM actions and response templates for therapeutic responses. Once the appropriate therapeutic response templates are retrieved for a step, they are filled with information from the conversational state tracker, e.g., *name, time of day, emotion*, *symptom*, *goal*, or *problem*.

   | | |
   |---|---|
   | Utterance: | I kept thinking about my child having an asthma attack. |
   | … | |
   | Template: | Earlier you mentioned that you were [symptom] |
   | Response: | Earlier you mentioned that you were worried. |

After two steps that solicit open-ended responses from the client, the utterance and inferred emotional state of the client are additionally passed to a separate model which predicts a ranked list of empathetic response from a list of 78 responses crafted toward caregiver experiences collected through a separate diary study. In the control condition, the responses were provided as a randomly sorted list. For the other turns, simple empathetic responses were shown and not used in the evaluation as there was no best choice[11].

*Study Recruitment*

The study team sent invitation emails to listservs (e.g., university programs students and trainee listservs) and applied snowball sampling. Inclusion criteria included: self-described English proficiency and had internet access. Additional criteria for specific groups included: for the novice group, participants needed to be registered nurses *without* extensive mental health training or the experience of having worked in a behavioral/psych unit for more than six months; for the mental health provider group, participants needed to be at least a third year in the clinical psychology program or a resident in the psychiatry program. These criteria were checked using an intake survey and verified by study team members. Trainees in the psychiatry and clinical psychology programs were considered "expert" and nurses were considered "non-expert" in mental health care. In total, 29 care providers responded to the screening survey. Due to the data retention policy of the communications vendor, the conversational data was only stored for seven days and then automatically deleted for nine participants, unbeknownst to the research team. After discovery of this limitation, an offline backup procedure was put in place to preserve the data of the remaining twenty participants. Of the 20, 11 were clinical psychologists or psychiatrists, and nine were nurses without extensive mental health care experience.

*Study Procedure*

A 2x2 study design was used to compare performance between mental health providers and nurses (2 groups) in delivering PST, whereby each participant completed one session without AI-enhanced response suggestions and one with AI-enhanced response suggestions. The order of these two conditions was randomized to control for practice effects in the evaluation. Each session was treated as a new scenario delivering a first session of PST.

Participants were invited to join an online session via video telecommunications software. The moderator followed a script to have the participant watch a tutorial video and then complete two sessions. Each session consisted of an exchange between the care provider and a study team member acting as a standardized patient to portray a family caregiver persona. The two scenarios presented different care symptoms (i.e. stress and sleep disturbance) with the order randomly assigned. After the two sessions, a study team member conducted an interview asking them about their impressions and suggestions for the overall system, specific features (e.g., empathetic responses, therapeutic responses). Participants were asked if the interview could be recorded, based on their responses the answers to the questions were either transcribed through a transcription service or taken as notes by the moderator or a notetaker. The participants also completed a survey of the System Usability Scale (SUS)[12].

*Provider Response Evaluation Criteria*

During certain steps as part of the therapeutic response, the operator must identify the symptom that the caregiver is experiencing and indicate this understanding to the caregiver. Additionally, within each session, there were two open-ended prompts to which the standardized patient provided descriptions of events impacting their emotional state (Figure 3). In response to these messages, the participants were presented with an extended list of empathetic responses and requested to select the correct simple reflection. For the remainder of the turns, A2P2 suggested simple affirmations, e.g. "Got it.", which were less context-dependent and thus interchangeable, so were not evaluated for accuracy or speed. The gold label responses for both scenarios are provided below:

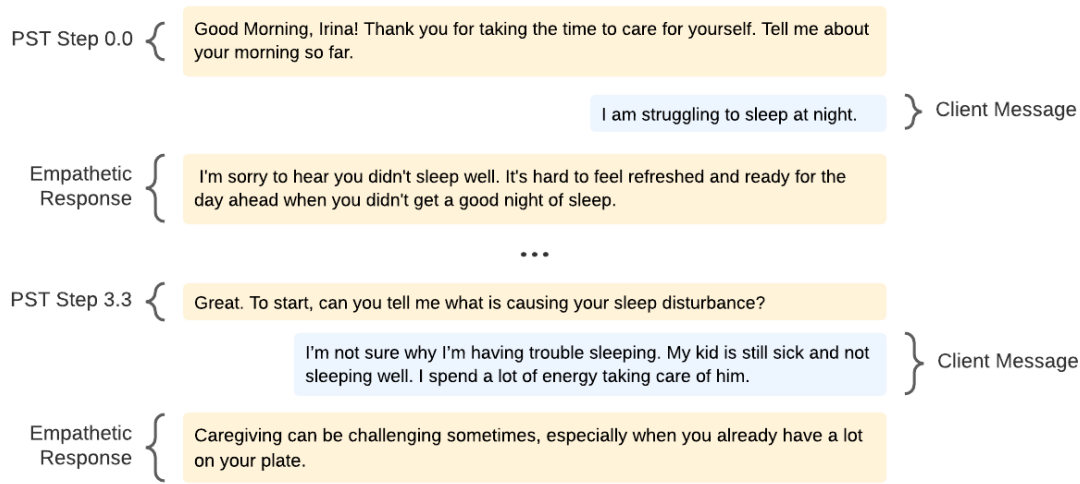

**Figure 3:** Excerpts from the sleep disturbance session showing the two opportunities for empathic responses.

In the control group, participants were asked to select *goals* for the client based on their symptoms (Figure 4). They were presented with five *goal* options, a pair for each *symptom* from the knowledge graph that are exclusive to that *symptom*. In the intervention condition, the system provided only one set of *goals* based on the *symptom* detected by the natural language understanding model and stored in the conversational state tracker.

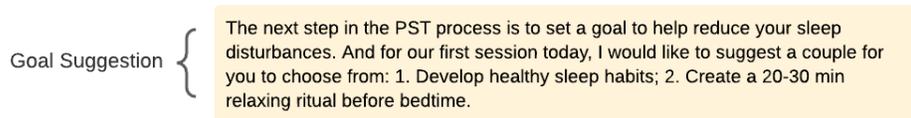

**Figure 4:** Goal suggestion for sleep disturbance.

*Data Analysis*
We hypothesized that care providers would make: **H1**) *faster responses* when presented with a ranked list of responses than a random list that contains the correct answer; **H2**) *faster* and *more accurate empathic responses* when presented with a ranked list of empathic responses than a random list that contains the correct answer; **H3**) *more accurate goal recommendations* when presented with goals that match the client's goal than when presented with all goal choices; and **H4**) *more accurate predictions of a caregiver's symptoms* when presented with a ranked list of symptoms than with an unranked list of symptoms. Each hypothesis was evaluated using permutation testing (with the exception of **H3**, which used a Fisher's Exact Test due to the small sample size and discrete values ranging from 0 to 2) to compare either the intervention and control conditions or the group differences between mental health providers and nurses. The effect sizes between groups and conditions were calculated using Cohen's D.

To usability and feasibility, the transcripts and notes from the exit interviews were coded by the lead author through deductive content analysis[13] based on a pre-defined list of categories of interest. The codes were iteratively refined through peer debriefing to arrive at the final set. We used descriptive statistics to examine the SUS total score.

**Results**

The non-expert group (n=9) was composed of a majority of nurses with more than five years of experience (n=6), two nurses with 3-5 years of experience, and one nurse with less than one year of experience working directly with patients and families. Clinical psychologists and psychiatrists also tended to be more experienced in their field with the majority (n=8) in their third year of residency or beyond.

*Aim 1: Preliminary Efficacy of the A2P2*

*Reduced response times (H1).* Response times were measured through the interface and used to evaluate **H1**. The response times averaged across all conversation turns in the session are normally distributed across participants in both the control and intervention groups (Figure 5). A permutation test was used to determine the statistical significance of the reduction (29.34%) in response time between the intervention and control conditions (Table 1).

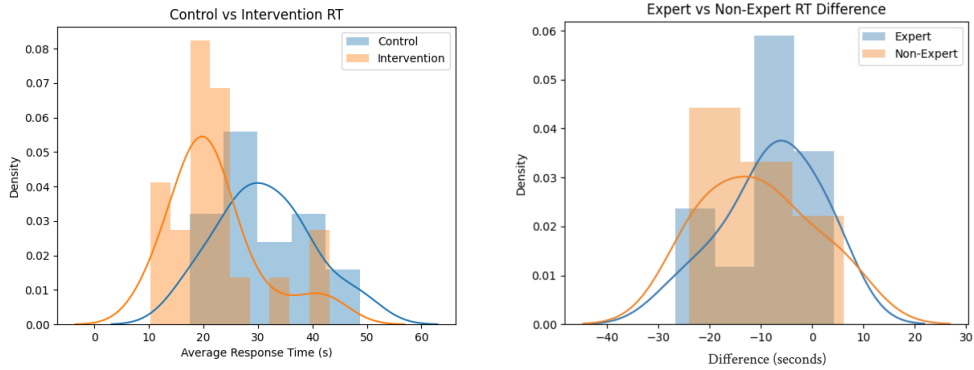

**Figure 5:** Probability distribution plot of the average response times in seconds (left) and the average difference in response times in seconds between the intervention and control for experts and non-experts (right).

The reduction in average response time (RT) between the intervention and control conditions was found to be statistically significant (p = 0.002) with a large effect size (d = 1.08). The conversational assistance system reduced average response times by 29.34%.

**Table 1:** Permutation test results for average response times per message across all participants in the non-expert, expert, and combined groups.

|  | All (n=20) | Non-Expert (n=9) | Expert (n=11) |
|---|---|---|---|
| Intervention Average RT (s) | 22.089 | 21.55 | 22.53 |
| Control Average RT (s) | 31.26 | 32.15 | 30.54 |
| Reduction | 29.34% | 32.97% | 26.22% |
| p-value | 0.002 | 0.027 | 0.049 |
| Cohen's D | 1.08 | 1.21 | 0.92 |

As shown in Table 1, non-experts had a greater mean reduction in response time than experts to the point of surpassing expert response times. However, this difference in relative reduction between the two groups was not significant (p=0.577).

The response time for the two turns, which included high-empathy responses, were compared between the intervention and control conditions to evaluate the speed portion of **H2**. A permutation test was used to determine the statistical significance of the decrease in response time (23.09%) between the intervention and control. The reduction in response time was not found to be statistically significant (p=0.082, d=0.4) between the intervention and control conditions. This may be due in part to the operators quickly selecting simple responses such as "I'm sorry to hear that" and not searching for higher-empathy responses. On average, the novice group decreased their response times by 29.46%, while the expert group decreased their response times by 18.55%. However, the difference in relative reduction between these two groups was not statistically significant (p=0.751).

*Increased Empathy with no reduction in Empathetic Response Time (H2).* The number of correct empathetic response selections are recorded in the contingency table below (Table 2). Based on a Fisher's Exact Test, the association between the AI-intervention (intervention vs control) and number of correct empathic responses (0, 1, 2) as defined in **H2** was found to be statistically significant ($p<0.001$).

**Table 2:** Contingency Table for Empathic Response Accuracy.

|            | Control | Intervention |
|------------|---------|--------------|
| Zero correct | 12      | 2            |
| One correct  | 7       | 9            |
| Two correct  | 1       | 9            |

The most common empathic response suggestions are presented in Tables 3 and 4 to provide further context into the differences in performance between the intervention and control groups.

**Table 3:** Most common empathetic responses in the control group.

| Count | Empathic Response |
|-------|-------------------|
| 18    | I'm sorry to hear that. |
| 4     | I'm sorry to hear you haven't been sleeping well. It's hard to feel refreshed and ready for the day ahead when you didn't get a good night of sleep. |
| 3     | That sounds difficult. I'm sorry you had to go through that. |
| 3     | That sounds difficult. Kids bring us joy but sometimes can be hard to deal with. |
| 2     | Work can be very stressful and overwhelming. Taking care of ourselves is especially important right now. |

**Table 4:** Most common empathetic responses in the intervention group.

| Count | Empathic Response |
|-------|-------------------|
| 9     | That sounds difficult. Kids bring us joy but sometimes can be hard to deal with. |
| 8     | I'm sorry to hear you haven't been sleeping well. It's hard to feel refreshed and ready for the day ahead when you didn't get a good night of sleep. |
| 8     | Caregiving can be challenging sometimes, especially when you already have a lot on your plate. |
| 6     | Work can be very stressful and overwhelming. Taking care of ourselves is especially important right now. |
| 4     | Difficult conversations are unavoidable. And they can be pretty stressful sometimes. |

*Improved treatment accuracy of goal selection (H3).* In the control condition, the novice group selected the correct goals 66% of the time whereas the expert group selected the correct goals 54.5% of the time. Since the AI-Assisted provided the correct goal corresponding to the client symptom, this amounted to a 60% increase in goal selection accuracy across all participants with statistical significance (p=0.001, d=1.13) supporting **H3**. Moderator notes on this phenomenon point to experts having strong opinions against certain goals selected by the subject matter experts that created the knowledge graph, e.g. journaling. This indicates that care recommendations for providers are not one-size-fits-all and may require personalization to their preferences and style of therapy.

*Symptom identification was not improved (H4).* There was no variability between the intervention and control groups in their ability to identify the symptom experienced by the caregiver in each scenario. For this reason, this experiment failed to support **H4**.

### Aim 2: Usability and Feasibility of A2P2

Results from the system usability survey indicated that the provider platform had *excellent* usability with an average score of 79.5 (standard deviation =12.2)[11]. Qualitative interviews post-study revealed that participants liked the platform overall and the AI-enhanced features. Providers in both groups perceived that text-based treatment could provide a private channel for caregivers to communicate asynchronously whenever they are available, lowering barriers to access of mental health services. Time savings was seen as one of the largest benefits of the system. Shortening the time required for care providers to respond was seen as benefiting the patient by increasing the time available to tailor the message to the individual, and being able to increase the volume of clients responded to in a given time period. The status quo in some cases for clients is "not getting a response for a day or two later".

Trust in the relevance of system-suggested responses was indicated as a factor in the speed at which they were able to respond. This points to a link between model performance and efficiency improvement. Familiarity with the response bank was indicated as another factor in reducing the time required to respond to messages (reducing reading time and improving trust). A clinical psychologist who did not use text-based messaging with patients indicated a desire to use the software to increase email response speed, "especially to crises", and reported delaying responding to messages due to "having to think of the right thing to say." Other providers echoed this sentiment, including a nurse who shared: "I like giving us a foundation for expressing concerns or empathy. It helps us stay professional sometimes when emotions are high.".

One care provider shared feedback they had heard from patients that similar systems made them feel like they were not treated as an individual and the responses felt canned. This corresponded to care providers indicating a need for additional responses and shared a desire to personalize the response bank to include their own variations so the client would not feel as if they were talking with a bot: "If I was typing in my own responses, if it would save them for me so that there's a quicker way to do that. And again, that would just make me feel more natural." The participants indicated appreciation for the inclusion of the slots within the templated responses, e.g. "Good [time of day], [name]!" → "Good Morning, Irina!".

Further, the controllable PST step tracker feature was well received by care providers and seen as a way to enable those without familiarity with the protocol to deliver it with high fidelity. For example, a psychiatrist shared, **"And I liked, again, how there was the problem-solving therapy structure was built into it in a way that seemed to make sense for somebody who's never done problem-solving therapy to work through those.".**

Overall, participants thought the system would be a good addition to their current workflow, and it would be feasible to implement it in their work setting. Participants viewed the system as part of a larger framework, where the system could serve as an adjunct for primary care and support healthcare workers without specialized training. In their view, this would help to expand mental health services and enable 24/7 access. Participants listed many other use cases for this technology including follow-up for patients in "high-risk scenarios", pediatrics, and oncology care which are family-focused, and used by healthcare workers without extensive training to follow up with patients. Check-ins between visits and screenings were suggested by four participants. Participants mentioned screening at primary care and use by other care professionals that are not trained in providing therapy for example staff at a clinic or a warmline for people to text to talk when needed based on their work experience in addiction and recovery. For example, a clinical psychologist shared, "So, this feels like a phone coaching kind of alternative that you could probably even get someone who's not even your therapist to do." A clinical psychologist saw this intervention as useful in the early stages of therapy but underscored the importance of escalating intense cases to human-expert care as part of longer-term interventions.

**Discussion**

Quantitative results indicate that with AI support response times are reduced, and the accuracy of empathic responses and goal selections are increased. These improvements are mediated by the addition of the response suggestion feature in both mental health providers and nurses. Exit interviews validated these results, showing that care providers perceived the technology as improving their efficiency and accuracy, as well as being easy to use. Additionally, the results indicate that goal selection accuracy was improved through the AI system validating H3. However, there was no improvement in symptom identification (H4).

In the initial development of the platform for WOZ data collection[9], the hypothesis was that empathic response data could be efficiently collected by providers interacting with standardized patients. However, providers would often use generic phrases such as "sorry to hear that", which indicated a need to support providers in delivering empathic responses as well. The results from the provider platform evaluation indicate that without AI-intervention providers continued to use simple responses such as "I'm sorry to hear that" (the top response). Whereas, the AI-intervention increased the use of higher empathy responses (Table 4). These results are consistent with recent studies of AI-supported authorship of empathic responses, which show that the perceived empathy of responses from peers can be improved by integrating a rewriting platform[6]. Though it may seem counterintuitive to use AI to augment the empathy of human beings, the current work provides further evidence that this capability can be supported computationally. In addition it shows that AI support for empathy is effective in the context of response selection also.

Qualitative feedback provided several insights into provider perceptions of the technology that indicate future directions for the improvement and application of similar systems. Participants in the expert group supported the system being used by individuals without mental health training to complete check-ins or low-intensity sessions with clients. This recommendation is supported by the quantitative result that the quality gap was closed between experts and non-experts using the system in the AI condition. Both experts and non-experts reported learning during the process, indicating an opportunity for the system as a training tool. One nuance related to acceptability was that providers were concerned about sounding robotic and wanted the opportunity to personalize the responses available through the system to sound more like themselves. By reducing cognitive load and the time required to respond to messages, providers felt they would have more time to personalize their messages for the patient.

In the dynamic landscape of artificial intelligence research, there is an increasing shift towards the development and implementation of generative AI solutions. Despite this trend, it is crucial to acknowledge that retrieval-oriented methods still have a significant role to play. For example, retrieval-based methods can contribute to generative models by supplying examples from a response bank, in a process referred to as retrieval augmented generation. This process has the potential to enhance the performance of generative models by enabling them to better adapt to human expectations and reduce "hallucinations" a phenomenon where generative models make up information in their responses. This represents a hybrid approach to AI development, leveraging both the creativity of generative models and the grounded specificity of retrieval methods, thereby producing output that is not only more human-like but also more controllable and thus safer for healthcare applications.

**Limitations**

The present study used a set of prototype responses crafted by the study team for a bot-delivered intervention rather than being derived from natural therapy dialog. Consequently, some responses were perceived as robotic and impersonal. Future research could explore applying advanced generative AI techniques to produce more natural and personalized responses. Moreover, the extent to which clients perceive AI-suggested responses as empathetic remains unclear and warrants further investigation. Another limitation of the study is the small sample size, necessitating caution in interpreting the findings. The results should be considered preliminary, and additional evaluations with larger and more diverse samples are needed to confirm and expand upon these initial findings.

**Conclusion**

By developing and evaluating an AI-assisted provider platform, we have shown the potential for this technology to assist care providers in expressing empathy to their clients while reducing response times and increasing treatment accuracy of goal selection. This work represents a promising step towards addressing the supply-demand imbalance in mental health services by reducing response times and lowering the educational barrier to entry, increasing the number and efficiency of care providers who can deliver protocolized therapies.